\documentclass[1p,11pt,times]{elsarticle}

\usepackage{ecrc}
\usepackage{epsfig}

\usepackage[cmex10]{amsmath}
\usepackage{amssymb}
\usepackage{amsthm}
\usepackage{algorithmic}
\usepackage{array}     	
\usepackage{subfigure}
\usepackage[normalem]{ulem}  
\usepackage[linesnumbered, ruled, vlined]{algorithm2e}
\usepackage{verbatim}  	
\usepackage{tabularx}  	
\usepackage{multirow}  	
\usepackage{booktabs} 	
\usepackage[table]{xcolor}  
\usepackage[bookmarks]{hyperref}
\usepackage{url}

\jid{procs}

\usepackage{amssymb}

\usepackage[figuresright]{rotating}

\begin{document}

\begin{frontmatter}




\title{\bf An Improved E-voting scheme using Secret Sharing based Secure Multi-party Computation}


\author[rvt]{Divya G Nair\corref{cor1}}
\author[els]{Binu. V.P}
\author[els1]{G. Santhosh Kumar}

\cortext[cor1]{Corresponding author}

\address[rvt]{M Tech Student, Cochin University of Science and Technology. \\Email: divyagnr@gmail.com}
 
\address[els]{Research Scholar, Cochin University of Science and Technology.}  
\address[els1]{Assistant Professor, Cochin University of Science and Technology}

\begin{abstract}
E-voting systems (EVS)are having potential advantages over many existing voting schemes.Security, transparency ,accuracy and reliability are the major concern in these systems.EVS continues to grow as the technology advances.It is inexpensive and efficient as the resources become reusable.Fast and accurate computation of results with voter privacy is the added advantage.In the proposed system we make use of secret sharing technique and secure multi party computation(SMC) to achieve security and reliability.Secret sharing is an important technique used for SMC.Multi-party computation is typically accomplished using secret sharing by making shares of the input and manipulating the shares to compute a typical function of the input.The proposed system make use of bitwise representation of votes and only the shares are used for transmission and computation of result.Secure sum evaluation can be done with shares distributed using Shamir's secret sharing scheme.The scheme is hence secure and reliable and does not make any number theoretic assumptions for security.We also propose a unique method  which calculates the candidates individual votes keeping the anonymity.
\end{abstract}


\end{frontmatter}

\section{Introduction}
\label{section:Introduction}
The election process can be considered as an extremely critical task which functions as the key of democracy.So the security violation in any aspect will be a sensitive issue.The election system must be sufficiently robust to withstand a variety of malicious behaviors and must be sufficiently transparent and comprehensible that voters and candidates can accept the results of an election.We are exploring the field of secret sharing based SMC for the E-voting application where privacy needs to be ensured from different aspects.

In recent decades some electronic voting schemes have been proposed to make election systems more robust and accurate.The two most important properties in all of these schemes are privacy and universal verifiability.Privacy ensures that it must not be possible to track down the relation between ballots and voters (anonymity) and also voter must not be able to prove who he voted for (anti-coercion).Universal verifiability means that the correctness of election procedure must be clear for everyone.In other words it must be guaranteed that voters votes have been tallied in the same way that they have intended.

Since 1980, much work has been put on developing secure electronic voting schemes.The major design criteria was that the voting system should guarantee that voter should remain anonymous during the entire voting process and also it must be at least as secure as traditional voting systems.The EVS should also guarantee the privacy.That is the vote made cannot be traced back to the individual.Detailed description of the existing electronic voting protocols, cryptographic primitives used and their properties are given in \cite{fouard2007survey}.The main properties of a voting schemes are correctness, privacy,receipt-freeness, robustness, verifiability, democracy, fairness and efficiency.

It is known that none of the existing protocols satisfy all these properties at the same time.Most of the constructions  meet the major requirements depending on the type of the elections performed.The most commonly used cryptographic primitives in EVS systems are Interactive zero knowledge proofs, secret sharing techniques, homomorphic encryption, re encryption, blind signature, shuffling based schemes and secure multi-party computation.Secret sharing is used in electronic voting in order to get the voting protocol to be robust against authorities coercions.The proposed system make use of homomorphic properties of secret sharing scheme for vote tallying.It also provides secure computation since shares are used for the computation.

Development of secret sharing scheme started as a solution to the problem of safeguarding cryptographic keys by distributing the key among $n$ participants and $t$ or more of the participants can recover it by pooling their shares.Thus the authorized set is any subset of participants containing more than $t$ members.This scheme is denoted as $(t,n)$ \textit{threshold scheme}.The notion of a threshold secret sharing scheme is independently proposed by Shamir\cite{shamir1979share} and Blakley\cite{blakley1899safeguarding} in 1979.Since then much work has been put into the investigation of such schemes. Linear constructions were most efficient and widely used. A threshold secret sharing scheme is called \textit{perfect}, if less than $t$ shares give no information about the secret.Shamir's scheme is perfect while Blakley's scheme is non perfect.Both the Blakley's and the Shamir's constructions realize $t$-out-of-$n$ shared secret schemes.However,their constructions are fundamentally different. The proposed system make use of Shamir's secret sharing scheme.	
Shamir's scheme is based on polynomial interpolation over a finite field.It uses the fact that we can find a polynomial of degree $t-1$ given $t$ data points.A polynomial $f(x)=\sum_{i=0}^{t-1}a_ix^i$, with $a_0$ is set to the secret value and the coefficients $a_1$ to $a_{t-1}$ are assigned random values in the field, is used for secret sharing. The value $f(i)$ is given to the user $i$ as secret share. When $t$ out of $n$ users come together they can reconstruct the polynomial using Lagrange interpolation and hence obtain the secret. Shamir's scheme is perfect and ideal.The knowledge of $(t-1)$ pieces make secret data completely undetermined and also the share size is same as the secret. This scheme is easily computable when necessary data is available and it avoids single point of failure.Also it increases reliability, security, safety and convenience.

Secure Multi-party Computation \cite{goldwasser1997multi} is an active research area in cryptography.It allows a set of parties to compute a function of their inputs while preserving input privacy and correctness.It is often the case that mutually distrustful parties need to perform a joint computation but cannot afford to reveal their inputs to each other.This can occur, for example, during auctions, data mining, voting, negotiations and business analytics.The problem is how to conduct such a computation while preserving the privacy of the inputs.A secure multi-party computation problem deals with computing any probabilistic function on any input, in a distributed network where each participant holds one of the inputs, ensuring independence of the inputs,	correctness of the computation, and that no more information is revealed to a participant in the computation other than that participant's input and output.

SMC is dedicated to dealing with the problem of privacy-preserving cooperative computation among distrusted participants.It was first introduced by Yao in 1982 by putting forward the famous Millionaire's problem\cite{yao1982protocols}.This method is used to implement cooperative computation with all participants private data, ensuring the correctness of the computation as well as not disclosing additional information except the necessary results.If no participant in the multi-party computation can learn more from the description of the public function and the result of the global calculation other than what he can infer from his own information, the computation protocol is secure.    

SMC is accomplished here by secret sharing schemes.In secret sharing, the secret is not single handed, but multi-handed so that even if any of the parties involved in the computation are malicious, the secret can be reconstructed.A secret sharing scheme is verifiable if auxiliary information is included that allows parties to verify their shares as consistent.To handle malicious parties involved in any computation, the secret sharing scheme needs to be verifiable.  

We present a modified scheme which uses secret sharing based secure multi-party  for vote tallying.
Section 2 gives related work in this area.Section 3 and 4 explains the proposed system and algorithm.Experimental results are given in section 5.Security analysis of the scheme is mentioned in section 6.Section 7 is the conclusion.
\section{Related Work}
\label{section:related work}
The first electronic election scheme was proposed by David Chaum \cite{chaum1981untraceable} in 1981. Electronic voting systems, catering to different requirements, have been widely implemented and used.There have been several studies on using computer technologies to improve elections.In 1987 Benaloh \cite{benaloh1987verifiable} presents an election scheme based upon secret sharing and the prime residuosity assumption.Boyd et al \cite{boyd1990new} in 1990 proposed multiple key cipher without a trapdoor function and presents a voting scheme as an application of said cipher.Iverson and Kenneth \cite{iversen1992cryptographic} in 1992 made proposals of application of secret sharing technique and zero knowledge technique in secure election.Fujioka et al \cite{fujioka1993practical} suggested a  practical secret voting scheme for large scale elections in 1993.In this voting scheme voting is managed by an administrator who registers and authenticates voters and a counter who tallies votes. Benaloh et al \cite{benaloh1994receipt} also proposed a receipt free election scheme in in 1994.In 1997 Cranor et al \cite{cranor1997sensus} made an implementation of the Fujiyoka et al scheme.In 1998 a receipt free voting scheme for large scale election is proposed by Okamoto \cite{okamoto1998receipt}.Publicly verifiable secret sharing and its application to e-voting is proposed by Schoenmakers \cite{schoenmakers1999simple} in 1999.Lee and Kim \cite{lee2000receipt} proposed a modified scheme in 2000.Hirt et al \cite{hirt2000efficient} used homomorphic encryption scheme.Neff and Andrew \cite{neff2001verifiable} in 2001 suggested verifiable secret shuffle for e-voting.Malkhi et al \cite{malkhi2003voting} in 2003 gave constructions without cryptographic technique.The scheme uses secret sharing techniques and homomorphism.General secret sharing using Chinese remainder theorem with application to e-voting is proposed by Iftene \cite{iftene2007general} in 2007.Generalization of the Pailliar's crypto system and application to voting is proposed by Damagaard et al \cite{damgaard2010generalization} in 2010.Chen et al \cite{chen2014secure} suggested a scheme based on discrete logarithm problem and secret sharing in 2014.Enhanced scheme with more confedentiality and privacy is suggested by Pan et al \cite{pan2014enhanced} in 2014.

\section{Proposed System}
\label{section:proposed}
The above mentioned E-voting schemes work on different technologies used in the process of E-voting. The proposed system focuses the generation of secret shares, secure distribution of shares and secure computation of votes obtained for each candidate. The algorithm works on a bitwise-pattern representation of votes. 

The secrecy of vote is a sensitive issue which needs to be addressed with ultimate care. In the current Electronic Voting System, when a vote is casted, only that particular candidate's data(vote) is getting modified and it can be easily tracked. So in this paper we try to address this problem using SMC. Here when a vote is casted, rather than updating that particular candidate's data, the entire candidates data is getting updated which will make it difficult to track the vote.
\begin{figure}
	\centering
	\includegraphics[width=26pc, height=18pc]{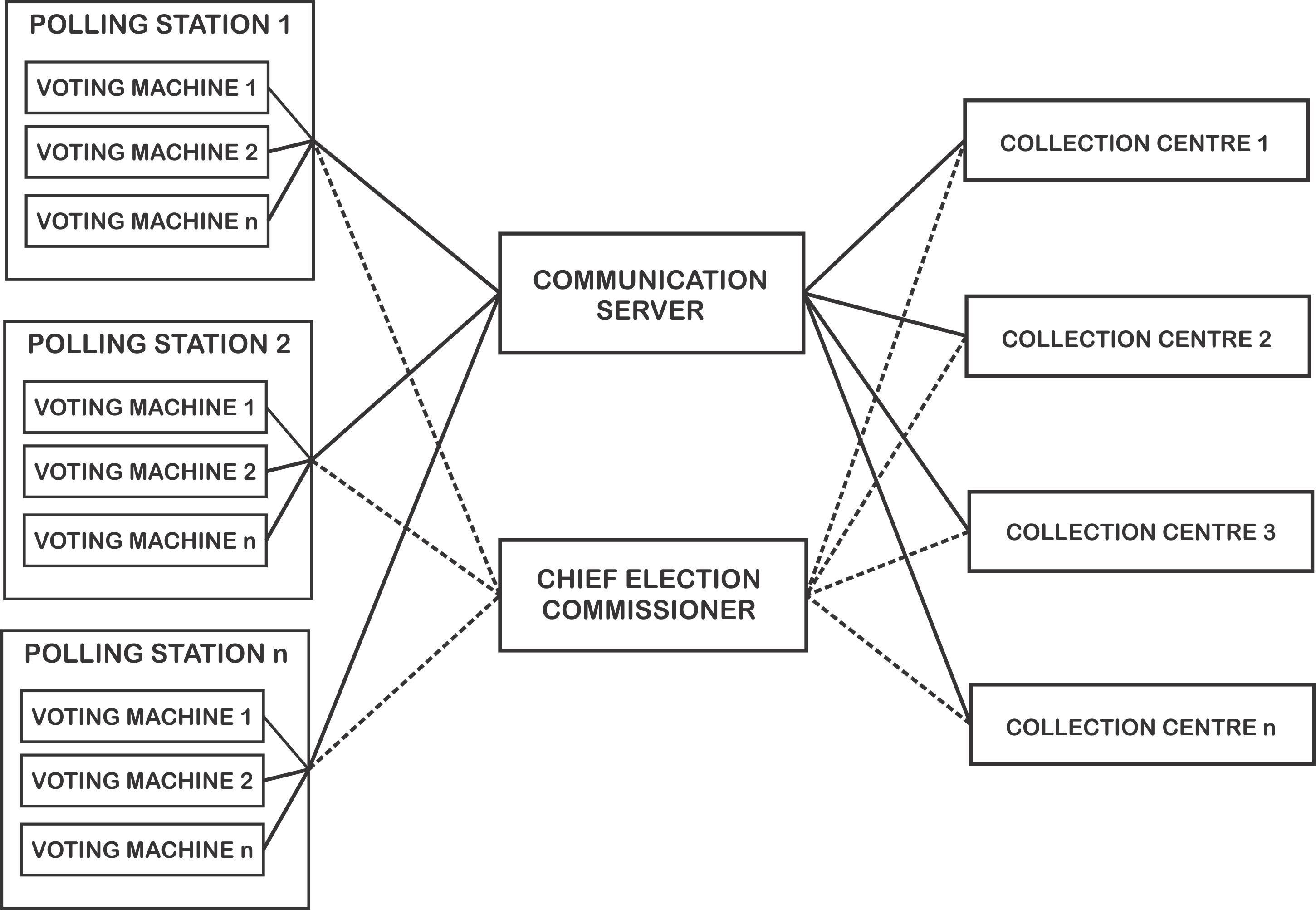}
	
	\caption{E-voting: System Architecture}
	\label{Arch}
\end{figure} 

In E-voting, each vote is considered as a secret and anonymization of voters identity should be protected. Each casted vote is divided into shares and these shares are distributed to multiple parties. The individual parties do not have any information about the vote casted by the voter by analyzing only the random share given to them.The individual shares are perfectly secure.The proposed algorithm also makes use of bitwise representation of votes which is distributed using Shamir's threshold secret sharing scheme and reconstructed using the coalition of specified number of parties.

\begin{figure}
	\centering
	\includegraphics[width=34pc, height=20pc]{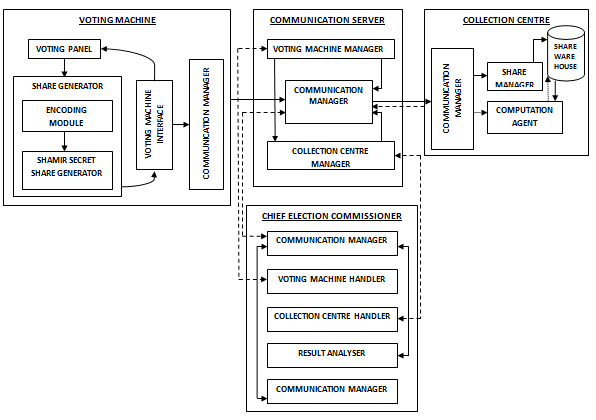}
	\caption{Detailed Architecture}
	\label{det}
\end{figure}

The voting system makes use of 4 important modules to accomplish the secure voting process.
\begin{itemize}
	\item Polling Station
	\item Communication Server
	\item Chief Election Commissioner
	\item Collection Centre
\end{itemize}
The System Architecture is shown in Figure \ref{Arch} and the detailed architecture of each module is shown in \ref{det}.	
Polling Station provides the interface for voting purpose. A polling station may contain many voting machines. It has a voting panel which contains the list of all candidates and their party symbols. Voting panel is loaded with this candidates information from a setup file  which is managed by the Chief Election Commissioner.The vote casted by a voter is given to share generator module which contains the encoding and Shamir share generator module.The Encoding module will emcode the vote using a bitwise encoding algorithm as explained in section \ref*{enc}.Share Generator uses Shamir's secret sharing scheme for generating shares of the encoded votes.The number of shares generated is based on number of collection centers.This provides both security and trust which is implemented using Shamir's $(k,n)$ threshold scheme in which any $k$ shares of total $n$ shares can be used for the reconstructing the original votes.
The shares generated in the share generator module is sent to the collection centres through the Communication Server which manages the communication and coordination among all the other  modules.This module handles Voting Machine Manager, Communication Manager and a Collection Centre Manager.Chief Election Commissioner module is working in an administrative role which manages the other modules. The Voting Machine Handler manages a set up file containing the list of candidates and their party symbols.Any modification made in the set up file will be reflected in the voting panel interface.The Collection Centre Handler manages the collection centres.For reconstructing the sum of votes for each candidate,$k$ collection centres need to be selected randomly based on $(k,n)$ threshold scheme.Collection Centre Handler randomly selects any of the $k$ collection centres while reconstruction.Authorization of Collection centres is also managed by Collection Centre Handler.The Result Analyzer analyses the whole Election Process and declare the result.

Collection Centre(CC) manages all the shares and provides a local database for holding the shares. Usually a group of authorized parties behave as collection centres.Each collection centre will be having a local database which receives one share for every vote casted.Number of collection centres($(n)$depend on the number of shares generated for each vote which in turn depends on the chosen threshold $(k,n)$ scheme.Each collection centre $CC_i$ gets $i^{th}$ shares of all the votes.By getting a share, collection centre does not have any idea regarding either the vote casted or the voter.The Computation Agent performs summation of all the shares it received in its local database and it is used as a partial sum in the multi party computation.When the Collection Centers are selected for the final result computation, the partial sum is passed to Collection Center Handler module in the Chief Election Commissioner module.The Result Analyzer compute the result  by reconstructing the encoded secret using Lagrange Interpolation in Shamir's scheme.The decoding algorithm is performed then, which will reveal the individual sum of votes for each candidate.

\section{Proposed Algorithm}
The algorithm for Share generation, distribution and result reconstruction is explained in Algorithm \ref*{Alg} EVS.
\label{sending side}
\begin{algorithm}
	\begin{scriptsize}
		\KwIn{Casted Vote}
		\KwOut {Individual votes obtained by each candidate}
		\BlankLine
		$m$=no\textunderscore of \textunderscore voters$()$\\
		$n$=no\textunderscore of \textunderscore candidates$()$\\
		Set the bit pattern and candidate bit block of sizes $n(1+lg m$) and $(1+lg m)$ respectively\\
		\For {each $voter$ $i$=$1$$:$ $m$  }{
			$encoded\textunderscore vote $ = set\_lsb (Candidate bit block) \\
			$V_i$ = bin\textunderscore decimal ($encoded\textunderscore vote $)\\
			Generate $k-1$ random numbers $r_1$,$r_2$,$r_3$ $\cdots r_{k-1}$\\
			Construct the polynomial  $f(x)=V_i+r_1x+r_2x^2+\cdots+r_{k-1}x^{k-1}$\\
			\For {$j$= $1$$:$$n$}{
				Generate share $X_{ij}=f(j)$, where $X_{ij}$ is the $j^{th}$ share of $i^{th}$ vote\\
				Send the share $X_{ij}$ to $CC_{j}^{th}$ collection centre through the secure communication manager\\
			}
		}   
		\For {each Collection Centre $j$ = $1$$:$$n$}{
			Sum of shares $SCC_j=\sum_{i=1}^{m}X_{ij}$ }
		\For {each randomly chosen Collection Centre $CC_k$}  {
			retrieve $SCC_k$ by the Collection Centre Handler} 
		Use the $k$ sum of shares $SCC_i$ ($i$:1 to $k$) interpolate and obtain the secret polynomial\\
		Decoding the constant term in the polynomial will give the individual candidates total votes 
		\caption{EVS}
		\label{Alg}
		
	\end{scriptsize}
\end{algorithm}
\subsection{Encoding and Decoding of Votes}
\label{enc}
The proposed system uses  bitwise representation of votes for the computation of sum. The number of bits required is based on the number of voters and also the number of contesting candidates.

The encoding process is explained below with an example.Consider we have 8 voters and 3 candidates.Each time a vote is casted, equivalent binary pattern will be generated by the encoder module.4 bits are required in this case for each candidate and a total of 12 bits since there are 8 voters and 3 candidates.The 12 bit vote pattern $b_{11}b_{10}b_{9}b_{8}$ $ b_7b_6b_5b_4$ $b_3b_2b_1b_0$  is initially set to $0$. When a voter votes for candidate 1, bit $b_{0}$ is set to 1, vote for candidate 2 is represented by setting bit $b_{4}$.similarly For candidate 3 bit $b_{8}$ bit is set.

The representation of vote is shown in Table \ref{table:voting process}.The share generator module will generate shares corresponds to the decimal equivalent of this binary pattern.Each time a vote is casted, a random polynomial is constructed and the shares corresponds to the secret vote is generated and sent to the Collection Centers.
For the  counting and result generation process, the Collection Center Manager receives $k$ sums of shares.The Result Handler then  apply Lagrange Interpolation formula to generate the polynomial $f(x)$ whose constant portion gives the sum of all votes. 
Decoding of this is then performed by taking each 4 bit combination of the binary pattern .The decimal equivalent of this is the candidates individual vote.

The Table \ref{table:share generation} gives the list of all shares generated for this example using a $(3,5$) scheme in which 5 shares are generated out of which any 3 can be used for reconstructing the secret sum and obtain the candidates individual sum of votes.

\begin{table}[ht]
	\small
	
	\caption{Voting System: Representation of votes}
	\centering
	\begin{tabular}{|l|c|c|c| } \hline
		{\bf Voter} & {\bf  Candidate} & {\bf  Representation} & {\bf  Secret}\\ \hline
		
		\hline
		
		Voter1 & Candidate1 & 0000 0000 0001 & 1\\
		Voter2 & Candidate3 & 0001 0000 0000 & 256\\
		Voter3 & Candidate1 & 0000 0000 0001 & 1\\
		Voter4 & Candidate2 & 0000 0001 0000 & 16\\
		Voter5 & Candidate1 & 0000 0000 0001 & 1\\
		\hline
	\end{tabular}
	\label{table:voting process}
\end{table}

\section{Experimental Results}
The implementation of the algorithm is done in java.The interfaces are developed with Netbeans IDE.Results based on 5 voters, 3 candidates and 5 Collection centers are considered.The shares  generated based on Shamir's $(3,5)$ scheme is shown  Table \ref{table:share generation}.CC1,CC2 and CC3 are chosen for the computation of the result.SCC1,SCC2,SCC4 obtained are.The Result Analyser uses these values for interpolation.The polynomial obtained is $275 +238x^1 +255x^2$. The constant term 275 represent the sum of votes.Decoding of this will result in 0001 0001 0011.Each 4 bit represent the individual votes obtained by candidates.

\begin{table}[!ht]
	\small
	\caption{Voting System: Share Generation}
	\centering
	\begin{tabular}{|c| c | c | c | c | c |}
		\hline 
		{\bf Collection Centre} & {\bf  CC1} & {\bf CC2} & {\bf  CC3} & {\bf CC4} & {\bf  CC5}\\ \hline
		
		
		Voter1 & (1,91) & (2,269) & (3,535) & (4,889) & (5,1331)\\
		Voter2 & (1,327) & (2,498) & (3,769) & (4,1140) & (5,1611)\\
		Voter3 & (1,70) & (2,251) & (3,544) & (4,949) & (5,1466)\\
		Voter4 & (1,113) & (2,278) & (3,511) & (4,812) & (5,1181)\\
		Voter5 & (1,167) & (2,475) & (3,925) & (4,1517) & (5,2251)\\
		
		\hline
	\end{tabular}
	\label{table:share generation}
\end{table}

\section{Security Analysis}
In traditional elections most ideal security
goals such as democracy, privacy, accuracy, fairness and verifiability, are assured to a certain level given physical and administrative premises.The task of meeting the security goals is quite difficult in online elections.Another controversial pair of security properties in E-voting schemes are privacy and eligibility.It is difficult in online elections to unequivocally identify and check the credentials of a voter, while at the same time protecting the privacy of his/her vote.Computerized voting will never be used for general elections unless there is a protocol that both maintains individual privacy and prevents cheating.

A good voting system should satisfy number of generic voting principle.The authentication mechanism should ensure that only eligible persons can vote and should not allow any one to vote more than once.The proposed method satisfies the fundamental requirement of a secure voting protocol.No one can determine for whom anyone else voted.Even the authorities will not be able to determine this because the information is not stored anywhere.For each vote casted the shares are send to all the collection centres and the partial sum is updated.The shares generated using Shamir's scheme is information theoretically secure and no information about the vote casted is obtained from the shares.The consistency of the result obtained can be verified with $k$ different set of shares.

\section{Conclusions}
\label{section:conclusions}
The voting scheme in this paper employs a new tallying method and helps to compute the individual candidates vote easily .It achieves the highest efficiency when the number of candidates
is small and guarantees strong vote privacy and reliability.We applied Shamir secret sharing scheme for Secure Computation of   E-voting results.The algorithm is promising and it ensures secrecy, integrity and efficiency in the process of electronic voting.The share generation and reconstruction are based on the bitwise representation of shares and the proposed method have several merits compared with existing cryptographic methods which are computationally complex.The bits required for the representation of votes is directly proportional to the number of candidates and voters.Each vote is encoded and random shares are generated which are information theoretically secure.We have not mentioned the authentication of users and collection center in this paper which will be managed by the Chief election commissioner module.The shares can also be sent to the collection center through different channels for ensuring more security.The research is still challenging
in the cryptographic community to design more powerful and secure schemes.

\end{document}